\documentclass[letterpaper]{article}
\usepackage{graphicx}
\usepackage{hyperref}

\begin{document}

\title
{\begin{flushleft}
       \mbox{\normalsize Published on: E. Journal of Theoretical Physics, \textbf{8}, 25 (2011) 343-352}
       \end{flushleft}
       \vskip 20pt
Calculating Vacuum Energy as a Possible Explanation of the Dark Energy}
\author{B. Pan
\footnote{email:
\href{mailto:bp6251@albany.edu}{bp6251@albany.edu}, or
\href{mailto:pan_b@sina.com}{ pan\_b@sina.com}.}
\footnote{Member of the $BaBar$ Collaboration at SLAC of the Stanford university since 2004.}\\
}
\maketitle

\begin{abstract}
We carried out a study of the properties of the $\lambda \phi^4$ field solutions. By constructing Gaussian wave packets to calculate the $S$ matrix, we show that the probability of the vacuum unbroken state transfers to the broken state is about $10^{-52}$. After adding this probability restriction condition as modulation factor in the summation of vacuum energy, we thus get a result that the vacuum energy density is about $10^{-47}GeV^4$, which is exact same as the observed dark energy density value, and maybe served as a possible explanation of the dark energy. Also our result shows that the vacuum energy density is proportional to the square of the universe's age, which fits the Dirac large numbers hypothesis.
\end{abstract}



\section{Cosmological Constant Problem}

In the current standard model of big bang cosmology, the $\Lambda CDM$ model, dark energy is a hypothetical form of energy in the space that accelerates the expansion of the universe. The most recent WMAP~\cite{wmap1}~\cite{wmap2} observations show that the universe is made up of about $74\%$ dark energy, $22\%$ dark matter, and $4\%$ ordinary matter. The dark energy is measured in the order of about $10^{-47}GeV^4$~\cite{pdg}. A possible source of the dark energy maybe the cosmological constant.

The Einstein's gravity field equation is
\begin{equation}
R_{\mu \nu} -\frac{1}{2}R\,g_{\mu \nu} + \Lambda\,g_{\mu \nu} = \frac{8 \pi G}{ c^4} T_{\mu \nu},
\label{eq:Einstein}
\end{equation}
in which $\Lambda$ is the cosmological constant. and it can also be written in the form of vacuum energy density
\begin{equation}
\rho_{\mathrm{vacuum}}c^2 = \frac{\Lambda c^4}{8 \pi G}.
\end{equation}

The general relativity field equation (\ref{eq:Einstein}) does not give any physical origin of the cosmological constant. We may explain it as the zero-point energy of the quantum fields. In quantum field theory, states can be treated as a set of harmonic oscillators which poses the zero-point energy
\begin{equation}
E=\sum_i\frac{1}{2}\hbar \omega_i ,
\end{equation}
which is obviously diverge when sum over all states. If we cut off the summation at the Planck energy scale
$E_{Plank} =\sqrt{ \frac{\hbar c^3}{G}} \simeq 10^{19} GeV$,
the zero-point energy density will be
\begin{equation}
\rho = \frac{E_{Plank}^4}{4 \pi ^2 \hbar ^3 c^5} \approx 10^{74}GeV^4,
\end{equation}
which is more than $10^{120}$ times larger than the measured value of dark energy density. To cancel almost, but not exactly, the quantum field theory faces a challenge.

In this article, we try to treat the Higgs field energy as the vacuum energy. First we study the static solution of the $\lambda \phi^4$ field equation. Then we figure out the probability of the vacuum unbroken state transfers to the vacuum broken state. Finally with probability modulation factor, we calculate the vacuum energy. Our result will be compared to the experimental data.

\section{Solutions Of the \texorpdfstring{$\lambda \phi^4$}{lambda-phi 4} Field}

The Lagrangian density of the $\lambda \phi^4$ field in the (1+3) dimensions is:
\begin{eqnarray}
\cal L &=& - \frac{1}{2} \partial_{\mu} \phi \, \partial_{\mu} \phi - \frac{\mu^2}{2} \phi^2 - \frac{\lambda}{4}\phi^4,
\label{equ:Lagrangian}
\end{eqnarray}
in which $\phi$ is the Higgs field, and $g_{\mu \nu}=(1,1,1,1)$. The Higgs boson has not been found yet in experiments, nor does the Standard Model predict its mass. Varies theories and experiments estimate its mass between $115 GeV$ to $180 GeV$~\cite{pdg}. In the Standard Model, with $\mu^2<0$, the Higgs boson mass is given by $m_H = \sqrt{-2\mu^2}=\sqrt{2\lambda}\,\upsilon$, where $\upsilon$ is the vacuum expectation value of the Higgs field. After spontaneous symmetry breaking, $\upsilon =\sqrt{\frac{-\mu^2}{\lambda}} =\sqrt{ \frac{1}{2G_F}}\approx 246 GeV$, fixed by the Fermi weak coupling constant $G_F$. The vacuum unbroken state has $\langle \phi \rangle = 0$, and the vacuum broken state has $\langle \phi \rangle = \upsilon$.

From the Lagrangian (\ref{equ:Lagrangian}), the equations of motion is
\begin{eqnarray}
  \nabla^2 \phi - \frac {\partial^2} {\partial t^2} \phi = \mu^2 \phi + \lambda \phi^3 ,
\label{equ:motion-phi}
\end{eqnarray}
which can also be written as
\begin{eqnarray}
 \nabla^2 \left( \frac {\phi}{\upsilon}\right) - \frac {\partial^2} {\partial t^2} \left(\frac {\phi}{\upsilon}\right)  = \frac {m_H^2}{2} \left( \left(\frac {\phi}{\upsilon}\right)^3 -\frac {\phi}{\upsilon} \right),
\label{equ:motion-phi-reduced}
\end{eqnarray}
in which $m_H$ is the Higgs mass, $\upsilon$ is the broken vacuum expected value.

In one dimension, we know equation (\ref{equ:motion-phi}) has static kink/anti-kink solutions
\begin{eqnarray}
\phi = \pm \upsilon \, \tanh(\frac{m_H}{2}x).
\label{equ:kink-1d}
\end{eqnarray}

Because $m_H \approx 115 \sim 180 GeV$, slope of (\ref{equ:kink-1d}) at $x=0$ is
\begin{eqnarray}
{\left. {\frac{d \phi(x)}{dx}} \right|_{x = 0}} = 2\upsilon \, m_H >>0 ,
\label{equ:slop}
\end{eqnarray}
we can replace the 1-dimension kink solution in (\ref{equ:kink-1d}) with a step function in our calculation without sacrifice accuracy
\begin{eqnarray}
\phi (x) = \left\{ {\begin{array}{*{20}{c}}
   {\upsilon, \,\,\, \mathrm{when} \,\,x > 0,}  \\
   { - \upsilon, \,\,\, \mathrm{when} \,\,x < 0,}  \\
\end{array} \,\,\,\,\, \Rightarrow } \,\,\,  \right. \phi ^2 (x) = \upsilon ^2 .
\label{equ:approx}
\end{eqnarray}

By study the properties of the Jacobi elliptic functions $sn$ and $cn$, we can get other static solutions of equ. (\ref{equ:motion-phi}) in 1-dimension as
\begin{eqnarray}
 \phi (x) &=& \pm \upsilon \, \coth (\frac{m_H}{2}x),
\label{equ:coth-1d}\\
 \phi (x) &=& \pm \sqrt{2} \, \upsilon \, \mathrm{sech} (i\frac{m_H}{\sqrt{2}}x) = \pm \sqrt{2} \, \upsilon \, \sec (\frac{m_H}{\sqrt{2}}x) = \frac{\pm \sqrt{2} \, \upsilon}{\cos (\frac{m_H}{\sqrt{2}}x)},
\label{equ:sech-1d}\\
 \phi (x) &=& \pm i \sqrt{2} \, \upsilon \, \mathrm{csch} (i\frac{m_H}{\sqrt{2}}x) = \pm \sqrt{2} \, \upsilon \, \csc (\frac{m_H}{\sqrt{2}}x) = \frac{\pm \sqrt{2} \, \upsilon}{\sin (\frac{m_H}{\sqrt{2}}x)}.
\label{equ:csch-1d}
\end{eqnarray}

If $\phi (x) = F(A x)$ is a solution of the equ. (\ref{equ:motion-phi}) in 1-dimension, in which $F()$ is a function such as listed above in (\ref{equ:kink-1d}), (\ref{equ:coth-1d}), (\ref{equ:sech-1d}), (\ref{equ:csch-1d}) and A is a constant, we can construct $(1+3)$ dimensions solutions of the equ. (\ref{equ:motion-phi}) as
\begin{eqnarray}
\phi ({x_\mu }) = F\left( {\frac{A}{\sqrt { a_\mu a_\mu }}(a_\mu x_\mu) } \right),
\label{equ:3d}
\end{eqnarray}
in which $a_\mu$ are arbitrary constants and automatically sum over $\mu$. For instance, it's easy to verify the $(1+3)$ dimensions $(it, x, y, z)$ kink-like solution for equ. (\ref{equ:motion-phi}) as
\begin{eqnarray}
\phi ({x_\mu }) = \pm \upsilon \, \tanh \left( {\frac{{{m_H}}}{{2\sqrt {a_0^2 + a_1^2 + a_2^2 + a_3^2} }}(i{a_0}t + {a_1}x + {a_2}y + {a_3}z)} \right).
\end{eqnarray}

\section{Unbroken And Broken States Of the Vacuum}

The Higgs mechanism requires the vacuum unbroken state has $\langle \phi \rangle = 0$, and the vacuum broken state has $\langle \phi \rangle = \upsilon$. Solutions (\ref{equ:sech-1d}) and (\ref{equ:csch-1d}) have domain walls, and can only take values above $+ \sqrt{2} \, \upsilon$ or lower than $- \sqrt{2} \, \upsilon$, which does not fit the $\langle \phi \rangle=0$ requirement of the unbroken vacuum state. So we will not use them in this calculation.

To the solutions (\ref{equ:kink-1d}) and (\ref{equ:coth-1d}), because of their anti-symmetry shape respect to $x_\mu = 0$, both solutions have the property of $\langle \phi \rangle = 0$. And because of their asymptotic behavior at infinity, if $x_\mu$ goes like $0 \rightarrow \infty \rightarrow 0$, then $\phi(x_\mu)$ goes like $0 \rightarrow \upsilon \rightarrow \infty$. Combination of the solutions (\ref{equ:kink-1d}), (\ref{equ:coth-1d}) and transform method (\ref{equ:3d}) can let the $\phi$ field get values from $-\infty$ to $\infty$, which satisfied our requirement. So we will use static 3-dimensions solution
\begin{eqnarray}
\phi (x, y, z) = \pm \upsilon \, \tanh \left( {\frac{{{m_H}}}{{2\sqrt {a_1^2 + a_2^2 + a_3^2} }}({a_1}x + {a_2}y + {a_3}z)} \right)
\label{equ:vac-ub}
\end{eqnarray}
as the initial static state wave function for the unbroken vacuum state. The relation $\phi ^2 \approx \upsilon ^2$ is still hold even in the 3-dimension case. That the positive or negative sign of the $\phi$ field does not matter, because we will only need $\phi ^2$ in the following integration. This approximation greatly reduces the calculation complexity.

To distinguish from the unbroken state $\phi$, we will use $\varphi$ for the spontaneous symmetry broken vacuum state. We can construct a static local Gaussian packet to represent it, which is centered at $\infty$ with a narrow width. Or practically to say, it is centered at point $b=(b_x, b_y, b_z)$, in which $(b_x, b_y, b_z)$ are the coordinates of point $b$.
\begin{eqnarray}
\varphi  = \upsilon {e^{\frac{{ - \Gamma _p^2\left( {{{(x - {b_x})}^2} + {{(y - {b_y})}^2} + {{(z - {b_z})}^2}} \right)}}{2}}}{e^{i\left( {{p_x}x + {p_y}y + {p_z}z} \right)}}
\label{equ:vac-br}
\end{eqnarray}
in which $\Gamma^{2}_{p}$ is the packet width in momentum space; $p_x, p_y, p_z$ are the momentum of the field; $\upsilon$ reflects the role that equ. (\ref{equ:kink-1d}) and (\ref{equ:coth-1d}) envelope on the Gaussian packet. If values of each of $(b_x, b_y, b_z)$ is very large, $\varphi$ will be very close to $\upsilon$. In fact, since the slope in (\ref{equ:slop}) is so large, point $b$ even does not need to be very far away from zero.

\section{Packet Width}

We now try to estimate the width of the Gaussian packet in
equ. (\ref{equ:vac-br}). By definition, vacuum is the original
point for measuring the energy. So for the unbroken vacuum
state, energy $E_{\mathrm{unbroken}} = 0$. From the
Lagrangian(\ref{equ:Lagrangian}), we read out the unbroken
vacuum state has $m^2 = \mu^2$, with $\mu^2<0$. From the
energy-momentum relation
\begin{eqnarray}
E^2 = p^2 + m^2 ,
\label{equ:E-p}
\end{eqnarray}
the momentum will be
\begin{eqnarray}
p_{\mathrm{unbroken}}^2 = E_{\mathrm{unbroken}}^2 - m^2 = -\mu^2 = \frac{m_{H}^2}{2} .
\label{equ:momentum}
\end{eqnarray}

For the vacuum state after spontaneous symmetry breaking, because we have moved the original point to $\langle \phi \rangle = \upsilon$, the energy will be changed. If we think momentum is conserved, $p_{\mathrm{unbroken}} = p_{\mathrm{broken}}$, and now the broken state has mass square as $m^2 = m_{H}^2 = -2 \mu^2$, then
\begin{eqnarray}
E_{\mathrm{broken}}^2 = p_{\mathrm{broken}}^2 + m_{H}^2 = \frac{3m_{H}^2}{2} .
\label{equ:E-broken}
\end{eqnarray}

From equ. (\ref{equ:E-p}), we have $EdE = pdp$. If we treat $dE$
and $dp$ approximately as $\Gamma_E$ and $\Gamma_p$ for the
width of energy and momentum respectively, as explained in detail in ref.~\cite{Akhmedov},
\begin{eqnarray}
 \Gamma_E E = \Gamma_p p,
\end{eqnarray}
which gives
\begin{eqnarray}
 \Gamma_p = \frac{E_{\mathrm{broken}}}{p_{\mathrm{broken}}}\Gamma_E = \sqrt{3} \Gamma_E.
\label{equ:gamma-p}
\end{eqnarray}

\begin{figure}[htp]
\centering
\includegraphics[totalheight=2.5in]{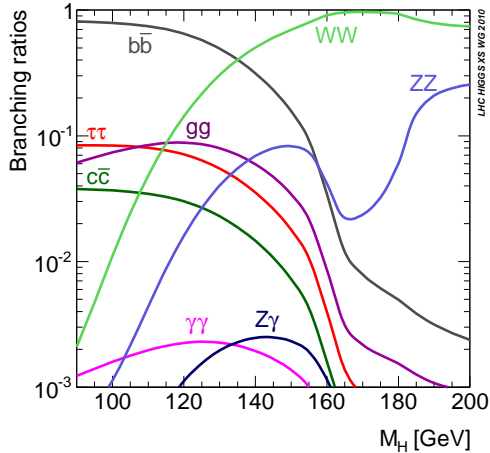}
\caption{List of most significant decay channels branch ratio. (Courtesy from CERN website~\cite{branch}).}
\label{fig:branch}
\end{figure}

\begin{figure}[htp]
\centering
\includegraphics[totalheight=2in]{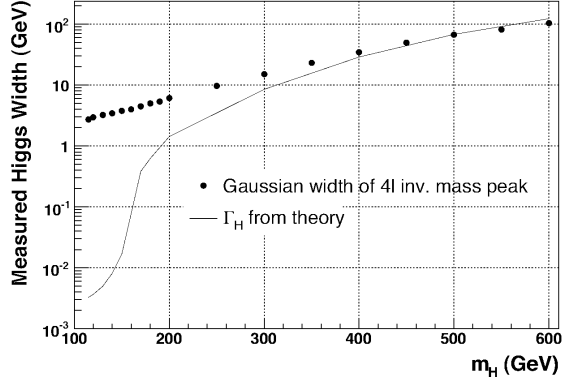}
\caption{Total Higgs width. The dots are simulation results from CMS (Physics TDR) for the $H \to ZZ^* \to 4$ leptons final state sub-threshold decay. The solid line is the Standard Model prediction.(Courtesy from Tully's talk~\cite{tully-talk}).}
\label{fig:width}
\end{figure}

Figure (\ref{fig:branch}) summarizes the branching fractions of the most important decay channels of the Higgs field. The Higgs decay width can be calculated in the Standard Model,
\begin{eqnarray}
 \Gamma (H \to f\bar f) &=& \frac{m_H}{8\pi} \left( {\frac{m_f}{\upsilon }} \right)^2 N_c \left( {1 - \frac{4m_f^2}{m_H^2}} \right)^{\frac{3}{2}},  \nonumber \\
 \Gamma (H \to WW) &=& \frac{m_H^3}{16\pi \upsilon ^2} \left( {1 - \frac{{4m_W^2}}{{m_H^2}}} \right)^{\frac{1}{2}} \left[ {1 - 4\left( {\frac{{m_W^2}}{{m_H^2}}} \right) + 12{\left( {\frac{4m_W^2}{m_H^2}} \right)}^2} \right],  \nonumber \\
 \Gamma (H \to ZZ) &=& \frac{m_H^3}{32\pi \upsilon ^2} \left( {1 - \frac{{4m_Z^2}}{{m_H^2}}} \right)^{\frac{1}{2}} \left[ {1 - 4\left( {\frac{{m_Z^2}}{{m_H^2}}} \right) + 12{{\left( {\frac{{4m_Z^2}}{{m_H^2}}} \right)}^2}} \right].
\label{equ:sm-width}
\end{eqnarray}
which is plotted in Fig. (\ref{fig:width}). When $m_H<140GeV$,
the dominant channels is the $H\rightarrow b \bar{b}$, and the
total decay width is in the $10MeV$ range.

The results in Fig. (\ref{fig:width}) is arguably for its
applicability. When $m_H$ is less than $2m_W$ or $2m_Z$, we may
research the 'sub-threshold decay', such as $H \to ZZ^*$, where
one Z boson is on-shell and the second Z boson's mass is
off-shell. Simulation from CMS (Physics TDR) for the $H \to
ZZ^* \to$4 leptons final state result is also shown in Fig. (\ref{fig:width})~\cite{tully-talk}~\cite{tully-cms}. According
to equ. (\ref{equ:sm-width}), $H \to WW$ channel's width should
be a little greater than $H \to ZZ$. But experimentally, it's
very hard to measure the energy and momentum of the neutrinos
in the $H \to WW \to$2 charged leptons + 2 neutrinos case; compares to
the easy be detected one in the $H \to ZZ \to$4 charged leptons
channel.

For the Higgs mass around $130GeV$, $ZZ^*$ channel sub-threshold decay has $\Gamma \simeq 2GeV$. If we think the $WW^*$ channel is about same, then the total width for $m_H=130 GeV$ is $\Gamma_E \approx 4 GeV$. That means, from (\ref{equ:gamma-p}), $ \Gamma_p = \sqrt{3} \Gamma_E \approx 6.8 GeV$ for $m_H=130 GeV$.

\section{Transition Probability Of the Vacuum States}

The lowest order $S$ matrix involves the transition amplitude of the $\phi^4$ interaction, which made of two incoming vacuum unbroken states $\phi _{i1} \phi _{i2}$ in equ. (\ref{equ:vac-ub}), and two out-going broken states $\varphi _{f1} \varphi _{f2}$ in equ. (\ref{equ:vac-br}),
\begin{eqnarray}
 S_{if} &=& \frac{4!}{2 \times 2} \cdot \frac{\lambda}{4} \,\, \langle \phi _{i1} \phi _{i2}|\varphi _{f1} \varphi _{f2} \rangle \nonumber \\
  &\approx & \frac{3 \lambda \upsilon ^2}{2} \int \varphi _{f1}\varphi _{f2} \, d^4 x \nonumber \\
  &= & \frac{3 \pi^{3/2} \lambda \upsilon ^4}{2 \Gamma _p^3} \,\, e^{ - \frac{p_x^2 + p_y^2 + p_z^2}{\Gamma _p^2}} \, \times \, \mathrm{phase} \, \times \, \mathrm{displacement} \times \int_0^{now} {dt} \nonumber \\
  &=&  \frac{3 \pi^{3/2} \lambda \upsilon ^4 T}{2 \Gamma _p^3} \,\, e^{ - \frac{p^2}{\Gamma _p^2}} \, \times \, \mathrm{phase} \, \times \, \mathrm{displacement} \nonumber \\
  &=& 10^{-26} \, \times \, \mathrm{phase},
\label{equ:amplitude}
\end{eqnarray}
in which the $phase$ and $displacement$ terms are
\begin{eqnarray}
\mathrm{phase} = e^{i(p_x (b_{1x}+b_{2x}) +p_y (b_{1y}+b_{2y}) + p_z (b_{1z}+b_{2z}))},
\end{eqnarray}
\begin{eqnarray}
\mathrm{displacement} = \exp \left[ -\frac{\Gamma _p^2}{4} \left((b_{1x}-b_{2x})^2 + (b_{1y}-b_{2y})^2 + (b_{1z}-b_{2z})^2 \right) \right].
\end{eqnarray}
The approximate sign in equ. (\ref{equ:amplitude}) means we replace the square of
$\tanh $ function with $\tanh ^2 \approx 1$ as in (\ref{equ:approx}). So only the Gaussian integrations of $\varphi$ are left. The phase term does not contribute to the following probability calculation. So we just omit it. The displacement term reflects the effect that the two out-going broken states centered at different points $b_1=(b_{1x},b_{1y},b_{1z})$ and $b_2=(b_{2x},b_{2y},b_{2z})$. We just let them to be at same points, so the displacement term reaches its maximum value, which is $1$. Other values we used are, $\upsilon=246GeV$,
$\lambda= m_H^2 / 2\upsilon^2 \approx 1/8$, $m_H = 130GeV$,
momentum square $p^2 = m_H^2 /2$ is in
equ. (\ref{equ:momentum}), $\Gamma _p = \sqrt{3}\Gamma_E \simeq
6.8 GeV$. The time integral gives out a term, $T$, which is the
age of the universe, equals to about $1.3 \times 10^{10}$years
$= 6.2 \times 10^{41} GeV^{-1}$. Then we get the probability
that the vacuum spontaneously transfer from unbroken states to
the broken states,
\begin{eqnarray}
 \mathrm{probability} = \left| S_{if} \right|^2  = 10^{-52}.
\label{equ:probability}
\end{eqnarray}

\section{Summation Over Possible States With Probability}

Vacuum does not like matter, in which matter has so many possible states that can oscillate in many frequencies. The vacuum energy can not contain contributions from all arbitrary wavelengths, except those physically existed states that are permitted by the Hamiltonian, e.g. To calculate the summation of states, we made the following assumptions:

1) The only contribution to the vacuum energy comes from the spontaneous symmetry breaking, the Higgs mechanism; which means we do not care any other possible sources. Vacuum has only two states: unbroken state and broken state, which will be the only two permitted states that appeared in the summation. Because the vacuum unbroken state has the energy $E=0$, so the only state left in the summation is the vacuum broken state.

2) The summation needs some modulation factor. For instance, to avoid diverge in the thermal radiation expression, in the derivation of the Planck's formula of blackbody radiation, we not only add up the energy of photons, but also times the Bose-Einstein distribution factor $1/(e^{\frac{h\nu}{kT}}-1)$. In this paper, we chose the probability of the transition from unbroken state to the broken state as the suppression factor.

Finally we sum the energy in phase space and get the vacuum energy density,
\begin{eqnarray}
\rho  = \frac{4\pi p^2 dp \cdot E}{{(2 \pi )}^3} \times \mathrm{probability}  = 10^{ - 47} GeV^4 ,
\label{equ:density}
\end{eqnarray}
in which the momentum and energy are in
equ. (\ref{equ:momentum}) and (\ref{equ:E-broken}), $\approx (1
\,\,\mathrm{to}\,\, 2) \times 10^2 GeV$; $dp \approx \Gamma_p$;
probability term is in equ. (\ref{equ:probability}). It is amazing
to see that our result in (\ref{equ:density}) is exact same as the value that
observed in the experiments.

\section{Discussion}
By adding certain conditions in the calculation of the vacuum energy, we get a same result to the observed value and maybe served as a possible explanation of the dark energy. Higgs has not been found yet in any experiment. The numerical estimation of the transition probability is sensitive to the input mass and width value. So we need further experimental data to consolidate out calculation.

When we integrate the $S$ matrix as transition amplitude, we use static wave functions. It maybe asked why we have $E$, $\Gamma _E$, $p$ and $\Gamma _p$, but only $e^{-\Gamma _p^2 x^2}$ and $e^{ipx}$ appeared in the calculation expression, while $e^{-\Gamma _E t}$ and $e^{-iEt}$ did not show up? A possible explanation is that $E$ is only occurred after symmetry breaking. Its value affects the decay after symmetry breaking, but not related to the unbroken state.

The Dirac large numbers hypothesis~\cite{large-number1}~\cite{large-number2}~\cite{large-number3}, made by Paul Dirac in 1937, proposed relations of the ratios between some fundamental physical constants. The hypothesis has two important consequences:

1) Matter should be continuously created with time. The mass of the universe is proportional to the square of the universe's age: $M \propto T^2$. According to Dirac, the continuous creation of mass maybe either of two models. Additive creation, assumes that matter is created uniformly throughout space. Or multiplicative creation, created of matter where matter already exists and proportion to the amount already existing.

2) The gravitational constant, G, is inversely proportional to the age of the universe: $G \propto 1/T$.

In our analysis, the transition amplitude (\ref{equ:amplitude}) proportions to $T$, means that the vacuum energy density in equ. (\ref{equ:density}) has the character of $\rho \propto T^2$, in which $T$ is the age of the universe. Since the dark energy is about $74\%$ of the universities mass, we can roughly to say that equ. (\ref{equ:density}) fits the consequence of the Dirac large numbers hypothesis, in case we did not count the expansion of the universe. If our calculation is correct, this relation will have a serious effect on the evolution of the universe.

\section{Remarks}

The 1-dimension kink/anti-kink solution (\ref{equ:kink-1d}) is printed in many books. But we had not seen anywhere lists solutions like (\ref{equ:coth-1d}), (\ref{equ:sech-1d}), (\ref{equ:csch-1d}) and (\ref{equ:3d}). For such a non-linear equation, we do not know any systematic method to solve it. So we just use the dummy way by trying after trying. We note here for further suggestion.

\end{document}